\documentstyle{article}
\newcommand{\hide}[1]{}
\begin{document}

\markboth{{\bf An angle to tackle the neutrinos
}} {{G. Rajasekaran}}

\begin{center}
{\large{\bf AN ANGLE TO TACKLE THE NEUTRINOS}}\\
\vskip0.5cm
{\bf G. Rajasekaran}
\vskip0.35cm
{\it Institute of Mathematical Sciences, Chennai 600113, India.\\
and\\
Chennai Mathematical Institute, Siruseri 603103, India\\
e-mail: graj@imsc.res.in}
\vskip0.35cm
\end{center}

\vspace{1.5cm}
\underline{Abstract} : A brief history of the discovery of
neutrino oscillations and neutrino mass is presented highlighting
the recent breakthrough in the determination of a crucial neutrino
parameter by the Daya Bay and RENO reactor experiments. The 
importance of this parameter in the context of one of the goals
of the India-based Neutrino Observatory (INO) project and also in 
advancing the frontier of neutrino physics is explained.\\

\newpage

Recently, an important discovery was made by two neutrino
experiments, one in China, called Daya Bay(1) and the other
in Korea, called RENO(2). Both measured the flux of
the antineutrinos at some distance from a complex of
powerful nuclear reactors in their respective countries.
The measurements showed that the antineutrinos oscillated,
thus allowing the determination of a fundamental parameter
of neutrino physics, the reactor angle.\\

To appreciate the importance of this experimental discovery
one must go back in time, a little bit.\\

\vspace{2mm}

\section{Early history and INO}

\vspace{2mm}

India was a pioneer in neutrino physics. The very first
detection of atmospheric neutrinos was made in the Kolar
Gold Field (KGF) mines in South India in 1965. These are
the neutrinos produced in the upper atmosphere by cosmic
rays and hence are called atmospheric neutrinos. The KGF
laboratory was closed in the 90's because of the closure
of the KGF mines.\\

It is the further study of these cosmic-ray produced neutrinos
that led the Japanese physicists to discover neutrino oscillations
and their leader M Koshiba to win the Nobel Prize in 2002. 
We in India missed the boat. Can we recover the lost initiative?
We can and we must. The India-based Neutrino Observatory (INO)
has been conceived with this objective in view (3).\\

INO Laboratory and INO Centre will come up in Theni District
and near Madurai city respectively, both in the southern part of India. 
In the underground (more correctly, under-mountain) laboratory
a gigantic 50-Kton magnetised particle detector will be erected to
study atmospheric neutrinos.\\

The Japanese group led by Koshiba discovered oscillations of
atmospheric neutrinos. But, before that, R Davis in USA had already
got evidence for neutrino oscillations in his pioneering
experiments on neutrinos from the Sun. However the oscillations of
the solar neutrinos got the clinching evidence only from the
subsequent experiment done with the heavy-water detector in
the Sudbury Neutrino Observatory in Canada. Davis shared
the Nobel Prize with Koshiba.\\

Although neutrino oscillation follows directly from quantum
mechanics, it leads to a result of profound consequence for
Physics and Astrophysics - the result that neutrinos have mass.
That is the importance of the discovery of neutrino oscillations.
Neutrino mass is the only concrete evidence we have for physics
beyond the Standard Model(SM) of high energy physics and hence is
expected to take us beyond known SM.\\

\vspace{2mm}

\section{The three angles}

\vspace{2mm}

Neutrino oscillations occur among three types of neutrinos
that are known to exist. The mixing among these three types
is governed by a 3 x 3 unitary matrix which can be specified
by three angles (like the three Eulerian angles required
to specify the 3 x 3 orthogonal matrix that describes the
rotation of a rigid body) and a certain number of phases.\\

The three angles can be called atmospheric, solar and
reactor angles since they respectively control the oscillations
of atmospheric, solar and reactor neutrinos. The atmospheric
and solar neutrino studies had determined the atmospheric and
solar angles as about 45 degrees and 30 degrees respectively more
than 10 years ago. Now the Daya Bay and RENO experiments have
determined the reactor angle to be about 9 degrees.\\

Again let us go back in history a little bit. During the exciting
period in the 90's when neutrino oscillations were discovered,
our group at IMSc,Chennai was one of the earliest to initiate a
comprehensive study of both solar and atmospheric neutrino
oscillations using the full mixing among the three types of
neutrinos (4,5). Others were using toy-models of mixing among
two types of neutrinos only to describe solar and atmospheric
neutrinos separately.\\

Since we were working with the complete three-neutrino framework,
we became the first to analyze the reactor neutrino data
that came in 1997 from the CHOOZ experiment in France. Analyzing
the data within this framework(6) we showed that the reactor angle
was smaller than 12 degrees and also showed that as a consequence
the solar and atmospheric oscillations became approximately
decoupled. This decoupling played a major role in all the
subsequent analyses of atmospheric and solar data helping to pin
down the parameters in these two sectors more easily.\\

The upper limit remained as our only information on the crucial
reactor angle for the last 15 years until it was determined
this year to be 9 degrees, not far away from the upper limit.
Now we are ready to explain the importance of this measurement.
There are two points: one in the context of INO and the other
in the context of matter-antimatter asymmetry.

\vspace{2mm}

\section{Neutrino masses and INO}

\vspace{2mm}

Although oscillations establish that the neutrinos are massive,
their actual mass cannot be determined by oscillation experiments;
only the mass differences (actually differences of squares of
masses) are determined. Calling the three neutrinos as 1, 2 and 3,
the 2-1 mass difference is determined by the solar neutrino
oscillations while the 3-2 mass difference is determined by the
atmospheric sector.\\

The mass-square differences so determined turn out to be very tiny:
0.00007 and 0.002 in units of electron Volt (eV) squared for 2-1 and
3-2 respectively. The sign of the 2-1 mass difference is determined
to be positive but the sign of the other is not determined.
So although neutrino 2 is heavier than 1, we do not know whether
3 is heavier than the 2-1 doublet or lighter.\\

A major discovery item in the agenda of the big magnetised
particle detector at INO is to resolve this ambiguity in the
sign of the 3-2 mass difference and thus determine
the actual mass-ordering of the neutrinos. A non-zero value for
the reactor angle is crucial for this discovery and that is the
importance of the reactor angle for INO. The rather large
measured value of this angle has enhanced the optimism of
the INO Collaboration. However in order to achieve this discovery,
the collaboration and its managers have to execute all the
components of the project according to strict time schedules.
There is no time to lose.\\

\vspace{2mm}

\section{Matter-antimatter asymmetry}

\vspace{2mm}

This is about the phases of the 3 x 3 unitary mixing matrix for
the neutrinos. Earlier we mentioned the three angles of
this unitary matrix. In contrast to the orthogonal rotation
matrix which is made of real numbers, the unitary matrix is
made of complex numbers and one can show that these complex
numbers lead to matter-antimatter asymmetry in elementary
particle interactions. Matter-antimatter asymmetry (also
called CP violation) was discovered experimentally in 1964
and is an important topic in high energy physics.\\

A complex number can be written as a real number multiplied
by a phase factor and this phase is therfore the signal for
matter-antimatter asymmetry. And it is this asymmetry which is
presumed to be responsible for the evolution of a original
matter-antimatter symmetric universe into the present-day
asymmetric universe that contains only matter and no
antimatter. So that is the cosmological importance of the
phases in the unitary mixing matrix.\\

The question is: apart from the three angles, how many
phases exist?\\

It was the simple remark of Kobayashi and Maskawa that
the dimension of the unitary matrix has to be atleast 3
for a phase to exist that won a Nobel Prize for them in
the year 2008. That was in the context of quarks and so for
the 3 x 3 unitary mixing matrix for quarks there exists
precisely one phase. For the neutrino case, there is some
difference. Ignoring this difference for the moment, there
is one phase in the case of neutrino mixing too.\\

However if the reactor angle were zero, then as already
mentioned the 3-neutrino problem would be reduced to two
uncoupled 2-neutrino problems described by two 2 x 2 matrices
which will not have any matter-antimatter-symmetry-violating
phase. Hence the importance of the non-zero reactor angle that
couples the solar and atmospheric sector into one 3-neutrino
problem. Now that the angle has been measured and
found to be 9 degrees, the door is open for measuring
the CP-violating phase. That will however require long-base-line
neutrino experiments, in which neutrinos produced by accelerators
in Japan, Europe or USA will travel through thousands of kilometers
inside the Earth and be detected and analysed. In the second phase
of the INO experiment, the magnetised detector will play the
role of this end-detector.\\
                                                                        
This will lead to an understanding of the contribution of
the matter-antimatter asymmetry in the neutino sector to
cosmological evolution.\\

Two more brief points must be added to make the neutrino picture
more complete.\\

\section{Towards a complete picture}

\vspace{2mm}

1. Although neutrinos are now known to be massive from the existence of
neutrino oscillations, we do not know the value of the masses since
only the differences in neutrino-mass-squares can be determined from the
oscillation phenomena. Nuclear beta decay experiments (in particular decay
of tritium) can give the absolute masses. So far it has only led to an
upper limit of 2.2 eV. Since the mass differences are very tiny as
already mentioned above, we see that all the three neutrino masses
are clustered around a mass level below 2.2 eV. This is almost a
million times smaller than the mass of the lightest massive elementary
particle known until the discovery of neutrino mass, namely electron of
mass 0.5 MeV.\\

2. It must be pointed out that the fundamental nature of the neutrino
is still not known, namely whether neutrino is its own antiparticle
or not. If neutrino is its own antiparticle, then the Kobayashi-Maskawa
counting of the number of phases is not valid for the neutrino
sector. It has to be augmented by two more phases. The question
whether neutrino is its own antiparticle can be answered only by the
neutrinoless double beta decay experiment which is therefore the most
important experiment in all of neutrino physics. This experiment also
is a part of the INO project.\\

\section{Summary}

\vspace{2mm}

The reactor angle whose upper bound was found 15 years ago in
Chennai was determined only this year by Daya Bay and RENO. The
rather large value of this angle gives strong impetus to INO
to pursue without delay its original goal of determining
the neutrino mass ordering and also to participate in
the long-base-line neutrino programmes aiming to fix the
matter-antimatter-symmetry-violating phase which is of
cosmological importance.\\

\newpage

\section{References}

\vspace{2mm}

(1) F P An et al (The Daya Bay Collaboration), www.arXiv.org:1203.1669
    [hep-ex]\\
(2) J K Ahn etal (RENO Collaboration),www.arXiv.org:1204.0626 [hep-ex]\\
(3) INO website: www.ino.tifr.res.in\\
(4) M Narayan, M V N Murthy, G Rajasekaran and S Uma Sankar, Phys Rev D
    53, 2809 (1996)\\
(5) M Narayan, G Rajasekaran and S Uma Sankar, Phys Rev D56, 437 (1997)\\
(6) M Narayan, G Rajasekaran and S Uma Sankar, Phys Rev D58,031301(1998)\\

\end{document}